\documentclass[12pt]{aastex}
\usepackage{emulateapj5}
\shorttitle{LMXBs in NGC 4472 II}
\shortauthors{Maccarone et al.}

\begin{document}
\def\simlt{\mathrel{\rlap{\lower 3pt\hbox{$\sim$}}
        \raise 2.0pt\hbox{$<$}}}
\def\simgt{\mathrel{\rlap{\lower 3pt\hbox{$\sim$}}
        \raise 2.0pt\hbox{$>$}}}


\title{The Low Mass X-ray Binary-Globular Cluster Connection II: NGC
4472 X-ray Source Properties and Source Catalogs \altaffilmark{1}}


\author{Thomas J. Maccarone\altaffilmark{2}, Arunav Kundu\altaffilmark{3} and Stephen E. Zepf\altaffilmark{3}}

\altaffiltext{1}{Based on observations made with the NASA/ESA Hubble
Space Telescope, obtained at the Space Telescope Science Institute,
which is operated by the Association of University for Research in
Astronomy, Inc., under NASA contract NAS 5-26555, and on on observations
made with the Chandra X-ray Observatory.}
\altaffiltext{2}{Astrophysics Sector, SISSA/ISAS, via Beirut 4, 34014
Trieste, Italy; email: maccarone@ap.sissa.it}
\altaffiltext{3}{Department of Physics and Astronomy, Michigan State
University, East Lansing MI, 48824; email: akundu, zepf@pa.msu.edu}

\begin{abstract}

We present the results of a Chandra/HST study of the point sources of
the Virgo cluster giant elliptical galaxy NGC 4472.  We identify 144
X-ray point sources outside the nuclear region, 72 of which are
located within the HST fields.  The optical data show 1102 sources, of
which 829 have colors consistent with being globular clusters (with
only 4 in the restricted central 10'' region).  Thirty matches are
found between the two lists - these are likely to be low mass X-ray
binaries associated with globular clusters, while forty-two of the
X-ray sources have no optical counterparts to $V\simlt25$ and
$I\simlt24$, indicating that they are likely to be predominantly low
mass X-ray binaries in the field star population with a small amount
of possible contamination from background active galactic nuclei.
Thus approximately 40\% of the X-ray sources are in globular clusters
and $\sim$4\% of the globular clusters contain X-ray sources.  There
is suggestive evidence that the X-ray sources located in blue globular
clusters may have harder X-ray spectra than those located in red
globular clusters.  No statistically significant differences are found
between the X-ray properties of the field sources and the X-ray
properties of the sources located in globular clusters.  This study,
along with our previous result from Paper I in this series on the
similarity of the spatial profile of the field LMXBs, globular cluster
LMXBs, and the globular clusters themselves suggest that a significant
fraction of the observed low mass X-ray binaries in the field may be
created in a globular cluster then ejected into the field by stellar
interactions; however, by comparing the results for NGC~4472 with
those in several other galaxies, we find tentative evidence for a
correlation between the globular cluster specific frequency and the
fraction of LMXBs in globular clusters, a correlation which would be
most easily explained if some of the field sources were generated {\it
in situ}.  We show that isolated accreting very massive black holes
are unlikely to be observable with current X-ray instrumentation and
that these sources hence do not contaminate the LMXB population.  We
discuss the possibility that several point sources near the nucleus
and aligned perpendicularly to the radio jet may indicate the presence
of a disk wind responsible for the low radiative efficiency observed
in the nucleus of this source.

\end{abstract}

\keywords{galaxies:general -- galaxies:individual(NGC 4472) --
galaxies:star clusters -- globular clusters:general -- X-rays:binaries
-- X-rays:galaxies}

\section{Introduction}

Early X-ray studies of elliptical and S0 galaxies showed that they
were significant sources of X-ray emission,with both a soft, kT$\sim1$
keV X-ray component and a harder X-ray component (fitting to a $\sim6$
keV bremstrahlung spectrum or a $\frac{dN}{dE}\propto$$E^{-1.7}$ power
law) with a flux roughly proportional to the optical luminosity
(Matsumoto et al. 1997; White, Sarazin \& Kulkarni 2002 and references
within).  The soft component is associated with hot virialized
interstellar gas while the hard component is associated with low mass
X-ray binaries - systems where a neutron star or black hole is
accreting gas from a Roche lobe overflowing companion star
(e.g. Forman, Jones \& Tucker 1985; Trinchieri \& Fabbiano 1985).
Since the advent of the Chandra X-ray Observatory, these X-ray
binaries have been resolved, with about half typically associated with
globular clusters, ranging from at least 70\% in NGC 1399 (Angelini,
Loewenstein \& Mushtozky 2001) to 40\% in NGC 4472 (Kundu, Maccarone
\& Zepf 2002 - hereinafter Paper I), and a lower limit of 20\% in NGC
4697 (Sarazin, Irwin \& Bregman 2000).

The lifetime of an X-ray source steadily accreting at the typical
$10^{37}$ ergs/second detection threshold for Chandra observations of
elliptical galaxies is no more than a few hundred million years (see
e.g. Bhattacharya 1995).  Thus for a low mass X-ray binary (LMXB) in
an elliptical galaxy to be observable with Chandra, it must either
have been recently formed or it must be an outbursting transient
system (Piro \& Bildsten 2002).  Most of the low mass X-ray binaries
in spiral galaxies are formed when a massive star in a tight binary
system undergoes a supernova explosion and leaves behind a compact
remnant in a close enough orbit for Roche lobe overflow from the
secondary star to occur.  In most elliptical galaxies, little or no
star formation has occurred for several billion years, so the only
means of LMXB formation are stellar interactions or the evolution off
the main sequence of a binary companion to a compact star.  In
contrast to low mass X-ray binaries, high mass X-ray binaries are
those systems where the accretion is driven by the compact object's
capture of stellar wind from a massive, secondary star (typically an O
star or an early B-type star).  Because the lifetimes of these
early-type stars are typically $\sim10^7$ years or less, high mass
X-ray binaries should only appear in regions with active star
formation, thus eliminating them as possible sources in low redshift
elliptical galaxies and old globular clusters (i.e. all globular
clusters except the very young ones seen in currently merging
galaxies).

Globular clusters have long been known to be fertile breeding grounds
for X-ray binaries.  About 10\% of the Milky Way's LMXBs are contained
within globular clusters (see e.g. Liu, van Paradijs \& van den Heuvel
2001 and references within) , despite the fact that less than 1\% of
the stellar mass of the Galaxy is contained in globular clusters.  The
dominant mechanism for forming new X-ray binaries outside of globular
clusters in spiral galaxies cannot work in elliptical
galaxies. Furthermore, elliptical galaxies have higher fractions of
their stellar mass in their globular clusters (see e.g. Ashman \& Zepf
1998; Rhode \& Zepf 2001).  Hence it is not surprising that the
well-observed elliptical galaxies all show much higher fractions of
their LMXBs in globular clusters than does the Milky Way.

In paper I of this series, we explored what properties make a globular
cluster most likely to contain an X-ray source.  The red (i.e. metal
rich) globular clusters are $\sim3$ times as likely as the blue
(i.e. metal poor) clusters to host an X-ray source. As an aside, we
note that elliptical galaxies have a larger fraction of their globular
cluster system in the red globular clusters, which may be an
additional reason for them to have a higher fraction of their low mass
X-ray binaries in globular clusters.  The luminosity function for the
X-ray sources associated with globular clusters is consistent with
being the same as the luminosity function for the sources where there
is no globular cluster counterpart.  Here we extend that work by
presenting the source lists of the globular clusters and the X-ray
sources, and show the associated optical parameters for each X-ray
source.  We discuss additional possible correlations between the
source parameters in greater detail than in Paper I.  We compare
summed X-ray spectra of field sources with globular cluster sources,
of sources in blue clusters with those in red clusters, and of sources
at different luminosities.  We show that very massive black holes
accreting from the interstellar medium cannot make up the X-ray
sources in NGC~4472 or most other elliptical galaxies.  We note the
similarities between the properties of the field and cluster LMXBs and
consider the relative merits of different scenarios for producing the
field LMXB populations.  We discuss the nuclear emission and the
possible existence of an equatorial outflow.

\section{Reduction Procedure}
\subsection{X-ray Data Reduction}

NGC~4472 is a giant elliptical galaxy, located at a distance of about
16 Mpc (e.g. Macri et al. 1999) in the Virgo cluster.  Chandra
observed NGC 4472 for 39588 seconds on June 12, 2000.  We examine only
the data from the ACIS-S3 detector.  We remove hot columns following
the guidelines on the ``ACIS Recipes:Clean the Data'' web page at Penn
State University
(http://www.astro.psu.edu/xray/acis/recipes/clean.html).  Because of
the large diffuse background in these observations (due to the X-ray
emission from the interstellar gas) no sources are detected with fewer
than 10 counts so the flaring pixels are likely to be of minimal
importance, but we nonetheless remove flaring pixels using the
Flagflare script written by T. Miyaji.

We create three images for source detection - a full band image from
0.5 to 8.0 keV, a soft band image from 0.5 to 2.0 keV, and a hard band
image from 2.0 to 8.0 keV.  While some source photons may be missed in
the 8.0-10.0 keV range, it has generally been found that this band
contributed far more noise than signal even in the ACIS-I detector
(see e.g. Baganoff 1999); the ACIS-S chips are more sensitive to soft
X-rays than the ACIS-I, and less sensitive to hard X-rays, so ignoring
the 8-10 keV band should be even more beneficial for these
observations.  We then run WAVDETECT from the CIAO 2.2 package using
wavelet scales from 1 to 16 pixels spaced by factors of 2.  We set a
false source probability detection threshold of $10^{-6}$, which
should yield an expectation value of slightly less than one false
source over the entire ACIS-S chip.

We extract spectra using the psextract script from the Chandra X-ray
Center.  We define the source regions to be the 3$\sigma$ ellipses
given by WAVDETECT and the background regions to be an elliptical
annulus centered on the source position with the outer major and minor
axes equal to twice the major and minor axes of the source region,
respectively, and the inner region defined by the outer boundaries of
the source region.  The spectra are binned so that at least 20 photons
from the source region (which includes some background photons) are in
each channel to allow reliable $\chi^2$ fitting, although the highest
and lowest energy bins may have less than 20 photons.  We have not
corrected the response matrix for the time dependent changes in the
quantum efficiency; however, this is unlikely to present a significant
problem since the data analyzed are from the first year of Chandra's
operation and channels below 0.5 keV (where the problems are most
severe) are ignored (L. David, CXC memo, available at
http://cxc.harvard.edu/cal/Acis/Cal\_prods/qeDeg/index.html).  This
effect should be less than 10\% and should affect only the spectral
region from 0.5 to about 0.7 keV, and hence should not change the
measured spectral indices by more than about 0.05, an amount smaller
than the measurement errors for all the fits presented in this paper.

We then fit the spectra in XSPEC 11.0 (Arnaud 1996) to find the
fluxes.  In all cases, we freeze the neutral hydrogen column to the
value $1.6\times 10^{20}$ cm$^{-2}$ (the Galactic value from the FTOOL
nH which uses the measurements of Dickey \& Lockman 1990).  We include
channels from 0.5 to 8.0 keV and fit absorbed power law spectra for
all the sources.  Where there are fewer than 4 channels from 0.5-8.0
keV for the spectral fit, we freeze the spectral index, $\Gamma$
(defined such that $\frac{dN}{dE}=E^{-\Gamma}$) to 2.0 and fit only
the normalization.  We note that the 2.0 spectral index is a bit
softer than that typically used for these sources, but since we do not
compare the spectral indices of individual sources in this paper, this
difference should not be too important.  This will result in the
source luminosity estimates being about 35\% lower than those assuming
$\Gamma=1.5$ or 20\% lower than those assuming $\Gamma=1.7$, yielding
systematic uncertainties of the same order as the uncertainties due to
the small photon numbers for these sources; this may cause the
luminosity functions we presented in Paper I to be systematically
steeper than those presented by other authors who assume a harder
spectrum for the sources with few counts, but should not affect the
conclusions of this paper, since in this work, the luminosity
functions of different subsamples of the sources are compared to one
another and are computed in a consistent manner.  In all but a few
cases, the fits give $\chi^2/\nu<2.0$, which generally represents an
acceptable fit for such a small number of photon bins.  Error bars are
not estimated for the poorly fitting sources (i.e. those with
$\chi^2/\nu>2$), as errors for poor spectral fits are of little
meaning.  We estimate the fluxes of the sources using the flux command
in XSPEC.  The absorbed (i.e. observed) fluxes are presented in the
tables.  We then compute the inferred (0.5-8.0) keV luminosity
assuming a distance of 16 Mpc (Macri et al. 1999) and isotropic
emission after correcting the fluxes for absorption, which is
typically about a 5\% effect.

\subsection{Optical Data Reduction and Astrometry}

Given the high density of GCs, subarcsecond relative astrometry
between the Chandra and HST images is essential to accurately
determine GC-LMXB overlaps and to eliminate false matches. The problem
is compounded by the fact that the absolute astrometry of the Chandra
image and each of the HST-WFPC2 frames may be inaccurate by 1-2$''$,
and very few Chandra sources are seen in the outer HST fields. We
adopted the following procedure to bootstrap all the images to the
USNO A2.0 (Monet et al. 1998) system: Using the IRAF TFINDER routine
and the positions of USNO A2.0 stars we first tied the large field of
view, ground-based image of NGC 4472 of Rhode \& Zepf (2001) to the
USNO A2.0 astrometric system.  We then used the positions of the
globular cluster candidates identified by Rhode \& Zepf (now in the
USNO 2.0 system) to obtain new astrometric plate solutions for each of
the four HST frames using TFINDER. The r.m.s. error of the astrometric
solution achieved for each HST image, and the relative astrometry of
the 4 WFPC2 frames is of the order of a hundredth of an arcsecond. In
order to bootstrap the optical and X-ray images we attempted to
isolate background galaxies/AGNs, but found too few to obtain reliable
relative astrometry. Therefore we decided to use the obvious GC-LMXB
matches themselves to recalculate the plate solution of the Chandra
image and bootstrap it to the USNO system. Note that though we
calculate the plate solution of the X-ray image using GC-LMXB matches,
we then use the X and Y positions of the X-ray detections along with
the plate solutions to calculate the "final'' RAs and decs of the
X-ray sources, which we then use to determine whether an X-ray source
matches a GC. We determined the reliability of the X-ray plate
solution by using different subsets of the optical GC positions to
create the X-ray plate solution e.g. only the ground-based GC
candidates in the outer regions of NGC 4472, only the HST identified
GCs in the inner regions etc. and found that in each case we
identified the same set of GC-LMXB matches. Thus we are confident that
we have achieved 0.3$''$ r.m.s. relative astrometric accuracy by the
bootstrapping procedure identified here. This is likely to be the best
optical to X-ray astrometric accuracy obtained to date.  All positions
reported in the tables are in the USNO A2.0 coordinates.

\section{Source Catalogs}
\subsection{X-ray catalogs}
We detect a total of 148 sources - 136 in the full band, 11 in the
soft band only, and 1 in the hard band only.  Sources are considered
to be matches if there is a positional offset between the different
bands of less than 1 arcsecond.  Three sources are within 8 arcseconds
of the center of the galaxy and appear to be associated either with a
weak active galactic nucleus or with brightness enhancements in the
hot interstellar gas - these sources are discussed in section 7.  One
additional source appears to be a spurious detection, as WAVDETECT
assigns it a count rate of 1.5 counts, and visual inspection fails to
find evidence of a source at that location.  Furthermore, the
background region around this source contains more photons than the
source region.  This source is flagged in the table, with all its
parameter values except the position listed as zeroes.  The remaining
144 sources should mostly be X-ray binaries associated with NGC 4472,
with about 10 likely to be foreground stars or background active
galactic nuclei (see Paper I and references within).

In Table 1, we list the X-ray ID number (with the sources sorted by
right ascension), IAU name, power law index, flux in the 0.5-2.0, flux
in the 2.0-8.0 band, the luminosities from 0.5-8.0 keV assuming the
power law model, a distance of 16 Mpc, and isotropic emission.  We
include flags for the ``good'' sources (i.e. those which appear to be
real and to be point sources upon visual inspection), whether the
source is within one or more of the HST fields of view, and the
optical ID numbers of the optical counterparts where such counterparts
exists.

\subsection{Optical catalogs}

Within the HST regions, we detect 1102 optical sources whose half
light radii are small enough to be globular cluster candidates, using
the source detection algorithm of Kundu \& Whitmore (2001) - see also
Paper I.  In Table 2, we list the ID numbers (sorted by right
ascension), $V$ band magnitude, $I$ band magnitude, $V-I$ color, and
half-light radii.  We also include flags which indicate whether the
source is within the Chandra field of view, and whether the source has
a typical globular cluster color (i.e. $0.8<V-I<1.4$).  There are 928
globular cluster candidates in the HST fields, 829 of which are also
within the Chandra fields.  Four of these sources are within the
central 8 arcseconds of the galaxy, so we attempt to match only 825
with the X-ray catalog.  For sources with X-ray counterparts, the
X-ray ID number is listed; for sources without, an ``N'' appears in
this column.

\subsection{The matching catalog}

We match sources between the X-ray and optical lists, ignoring sources
within 10 arcseconds of the center of NGC 4472, where the brightness
of the hot gas and the possible weak AGN activity makes the X-ray
detection procedure likely to find sources which are not X-ray
binaries.  We find 30 X-ray sources within 0.7 arcseconds of an
optical source with optical colors consistent with being globular
clusters.  Two additional sources show optical colors outside the
globular cluster color range and are likely to be either foreground or
background objects.  In table 3, we list the X-ray and optical
properties of the 30 globular clusters with X-ray sources - the X-ray
ID number, the IAU name, the best fitting power law spectral index,
the 0.5-2 keV and 2-8 keV fluxes, the luminosity, the $\chi^2$ and
number of degrees of freedom of the spectral fit, the optical catalog
ID number, the position of the optical source, the $V$ magnitude, the
$I$ magnitude, the $V-I$ color, and the half-light radius of the
optical counterpart.  Throughout the remainder of this paper, we will
define the ``globular cluster sources'' to be the members of the
matching catalog and the ``field sources'' to be only those sources
which do not have an optical counterpart but are within the HST field
of view.  Sources outside the regions observed by HST are not included
in discussions of the ``field sources.''

\section{X-ray spectral models}
As this work may be of some interest to researchers generally
unfamiliar with the detailed spectral properties of accreting black
holes and neutron stars, we include a very brief review of the topic.
It is known that the spectral form of the X-ray binary systems in the
Milky Way galaxy changes with luminosity as the sources enter
different spectral states.  There are two key continuum components to
X-ray spectra of black hole binaries in the 0.5-8.0 keV range - a
multi-temperature thermal component which dominates at the lower
energy end of the spectral range, and a power law tail which dominates
at the higher energy end of the range.  The thermal component is
generally thought to arise from a geometrically thin, optically thick
accretion disk where the disk's temperature varies as $R^{-3/4}$
(Shakura \& Sunyaev 1973), and the power law tail is generally thought
to arise from Compton upscattering of the disk photons in a cloud of
hot electrons (see e.g. Sunyaev \& Titarchuk 1980).

The three basic bright spectral states seen in black hole systems are,
in order of increasing luminosity, the low/hard state, characterized
by a $\Gamma\sim1.5$ power law, the high/soft state, characterized by
a disk blackbody component with $kT_{inner}\sim1-2$ keV and a weak
$\Gamma=2.5$ power law, and the very high state, which is similar to
the high/soft state, except that the power law component, rather than
the disk component dominates the bolometric luminosity of the very
high state.  For Galactic sources, the rapid variability properties
can also help in classifying the spectral state of a source, but
extragalactic X-ray sources are far too faint for these techniques to
be of use.  For a more detailed discussion of the spectral and
variability characteristics of sources in the different spectral
states, we refer the readers to Nowak (1995) and Tanaka \& Lewin
(1995), which discuss black hole binaries, and van der Klis (1995)
which includes some discussion of black hole systems but focuses more
heavily on neutron star systems.

Neutron stars also show different spectral states but they are
generally classified using a different nomenclature (see e.g. van der
Klis 1995).  The low magnetic field neutron stars have typically been
broken into two classes - the atoll sources and the Z-sources,
although this distinction has been blurred by recent observations (see
e.g. Muno, Remillard \& Chakrabarty 2002).  The high-magnetic field
neutron stars are predominantly accretion-powered pulsars, where the
accreted matter is funneled down the magnetic poles.  About 90\% of
X-ray pulsars in the Galaxy are found in high mass X-ray binaries, the
few which are in LMXBs all emit at well below $10^{37}$ ergs/sec (Liu
et al. 2001 and references within).  The atoll sources (or
``atoll-state'' sources) are all also much fainter than our detection
limits.  Thus, all the neutron star sources detected in these
observations (and most observations of elliptical galaxies made to
date) are likely to be Z sources.  The spectra of Z sources are rather
complicated, but typically are well fit by a model dominated by
thermal Comptonization in an optically thick ($\tau\sim10$),
relatively low temperature ($k_BT\sim3$keV) medium (see e.g. di Salvo
et al. 2001).

In this paper, we will fit three spectral models to the observed data.
Two of these models, the power law model and the thermal
bremsstrahlung model, are standard spectral models.  These two models
have been fit to X-ray spectral data since the earliest days of X-ray
spectroscopy because they often provide good parameterizations of the
data.  The power law model fits a normalization and a spectral index
$\Gamma$, defined such that $\frac{dN}{dE}\propto E^{-\Gamma}$.  The
bremsstrahlung model is parameterized in terms of a normalization and
a temperature, generally expressed in keV.  The third model we fit is
the disk blackbody model (Mitsuda et al. 1984), which models the
spectrum of a standard accretion disk.  It is parameterized in terms
of a temperature at the inner edge, generally expressed in keV and a
normalization.  While other models, such as thermal Comptonization
models, may provide a more physically motivated description of the
spectra, these models typically require too many physical parameters
to be fit given the relatively small number of photons observed with
Chandra.

\section{The X-ray Spectral Parameters}
\subsection{Comparison between cluster members and non-members}

While a substantial fraction of the LMXBs in NGC 4472 (and other
elliptical galaxies) are clearly not associated with globular
clusters, it is not yet clear that these sources were formed outside
globular clusters.  Stars may be ejected from globular clusters in
stellar collisions, and this effect is likely to be most important for
hard (i.e. tight) binary systems (see e.g. Phinney \& Sigurdsson
1991).  Additionally, globular clusters may be destroyed by
evaporation and tidal interactions with the galactic potential,
leaving behind whatever X-ray binaries they contained as field sources
(see e.g. Murali \& Weinberg 1997; Vesperini 2000).  If the formation
mechanism for the bulk of the field sources is different than the
formation mechanism for the globular cluster sources, then one might
expect to see a difference between the X-ray spectra of the field
sources and those of the globular cluster sources.  On the other hand,
the luminosity functions of the field and cluster sources have already
been shown to be consistent with one another (Paper I).  Since the
spectral properties of accreting compact objects are generally
well-tied to the accretion rate in Eddington units (see the discussion
in the previous subsection), one would only expect a difference in the
spectral properties if the field and cluster sources if the sources
below the Eddington limit for a neutron star were composed mostly of
high Eddington fraction neutron stars in one case and of low Eddington
fraction black holes in the other.  It is not clear whether the field
and cluster LMXB populations should be contain the same relative
fractions of black hole and neutron star primaries.  Thus it may be
beneficial to test whether the spectral properties of the field and
cluster sources are consistent with one another.

Given this motivation, we extract a composite X-ray spectrum for the
field sources and composite spectrum for the globular cluster sources.
As above, we bin the data so that at least 20 photons are included in
each channel, ignore channels below 0.5 keV and above 8.0 keV, and fix
the neutral hydrogen absorption column at 1.6$\times10^{20}$cm$^{-2}$
(the Galactic value along the line of sight to NGC 4472).  We then fit
an absorbed power law model to both summed spectra.  For the field
sources, we find the photon index, $\Gamma$=1.44$\pm0.10$ (90\%
confidence interval), while for the globular cluster sources,
$\Gamma=1.38\pm0.10$, with values of $\chi^2/\nu$ of 66/71 and 45/65,
respectively.  A bremsstrahlung model gives
$kT_{GC}=17.5\pm^{20.5}_{6.5}$ keV and
$kT_{field}=11.5\pm^{10.3}_{4.1}$ keV, with values of $\chi^2/\nu$ of
44/65 and 63/71, respectively, while a disk blackbody model (diskbb in
XSPEC - see Mitsuda et al. 1984) gives $kT_{GC}=1.53\pm^{0.24}_{0.20}$
keV and $kT_{field}=1.38\pm^{0.22}_{0.17}$ keV for the inner disk
temperatures, with values of $\chi^2/\nu$ of 58/65 and 71/71,
respectively.  Thus we find that the summed spectra of the globular
cluster and field sources are consistent with being the same, and that
all three models provide statistcally acceptable fits for both the
field and the globular cluster integrated source spectra.

We note that the spectral indices and bremstrahlung temperatures we
find here indicate a somewhat harder spectrum than that typically
found for the hard excess in elliptical galaxies from ASCA
observations.  Multiple explanations are possible.  The first is that,
because our detection threshold is at $\sim$ a few times $10^{37}$
ergs/sec (the faintest source measured is at $10^{37}$ ergs/sec, but
there are severe incompleteness effects at this luminosity), we are
biased towards sources that have hard spectra in the 0.5-8 keV range.
To those familiar with state transitions in the X-ray band, this may
seem counter-intuitive, as the ``hard'' state for black holes
typically occurs at luminosities less than $\sim5\times10^{37}$
ergs/sec, so only a small fraction of the black holes observed should
be hard state objects.  However, the spectra of the high/soft state
are typically dominated by a thermal disk component with a temperature
of $\sim 1-2$ keV (i.e. similar to the observed composite spectrum).
The effective area of Chandra drops above the peak of the disk
blackbody, so moderate depth Chandra observations will not be
particularly sensitive to this cutoff.  Hence a power-law model fit,
dominated by the low energy portion of the disk blackbody spectrum, is
acceptable in $\chi^2$ terms, and will tend to have a flatter
(i.e. harder) spectral index than the typical low/hard state in the
0.5-8.0 keV range.  The ASCA observations, on the other hand, see a
larger fraction of their total flux from the low luminosity sources
which have true power law spectra (since they include all the flux
from X-ray binaries and not just the flux from those X-ray binaries
bright enough to be resolved in the Chandra observations).
Furthermore, the effective area of Chandra falls off more sharply
above 4 keV than does that of ASCA, so the best power law fit to a
$\sim1$ keV disk blackbody spectrum from Chandra will have a harder
spectral index than the best fitting power law model made by ASCA to
the same intrinsic spectrum (since ASCA will give more weight to the
low flux data above the blackbody's cutoff energy).

\subsection{Correlation between X-ray luminosity and spectral properties}
For the sources which have luminosities higher than the Eddington
limit for neutron stars, we attempt to determine if evidence for state
transitions between the different black hole spectral states can be
seen.  Below the Eddington limit for accreting neutron stars, it will
be impossible to disentangle, solely on the basis of luminosity,
neutron stars accreting at a high fraction of the Eddington luminosity
from black holes accreting at a lower fraction of the Eddington
luminosity.  We thus restrict this analysis to sources above the
Eddington limit for a neutron star.  The Eddington limit for a neutron
star occurs at about 15\% of the Eddington luminosity for a 10
$M_\odot$ black hole.  Typical luminosities for the state transition
from the low/hard to the high/soft state are $\sim5-10$\% of the
Eddington luminosity, while the transition from the high/soft to the
very high state typically occurs at about 30\% of the Eddington
luminosity, although hysteresis is sometimes seen in the state
transitions (see e.g. Nowak, Wilms \& Dove 2002; Maccarone \& Coppi
2002).

We create two groups of sources from which to make summed spectra -
the sources with luminosities between 2 and 4$\times10^{38}$
ergs/second, which are assumed to be in the high/soft state and the
sources with luminosities greater than 4$\times10^{38}$, which are
assumed to be in the very high state.  We include sources with and
without optical counterparts.  For a power law model, the high state
and very high state spectral indices are $1.42\pm0.09$ and
$1.44\pm0.08$ respectively, with $\chi^2/\nu$ of 51/55 and 74/66,
respectively.  For the disk blackbody, the high/soft state fits well
($\chi^2/\nu$=38/55) for $kT=1.32\pm^{0.17}_{0.14}$ keV, while the
very high state has a best fitting temperature of
$1.30\pm^{0.16}_{0.14}$ keV; however, the very high state spectrum
shows a residual excess above 4 keV and has $\chi^2/\nu$=95.8/66,
ruling out the model at the 99\% confidence level.  The summed
``high/soft state'' and ``very high state'' spectra are plotted in
Figure 1.

\begin{figure*}
\plotone{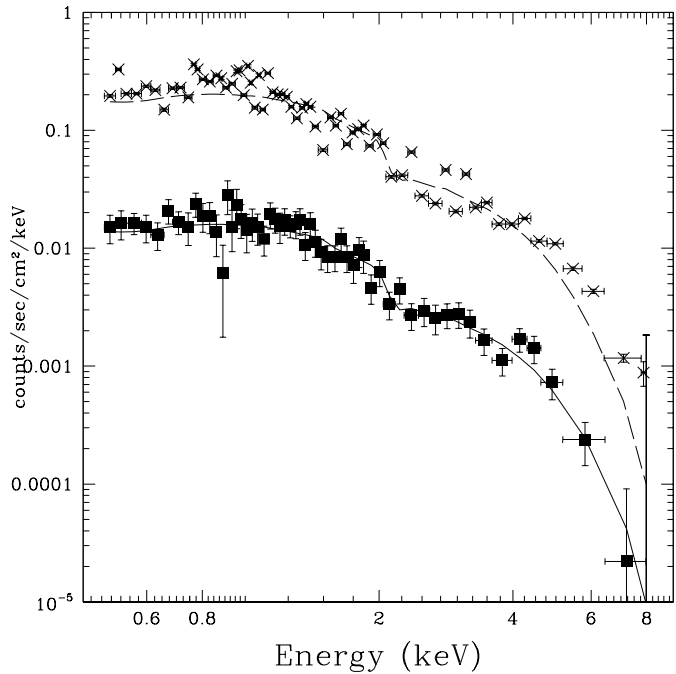}
\caption{The ``high/soft state'' integrated spectrum (filled squares)
and the ``very high state'' integrated spectrum (X's). The very high
state spectrum has been shifted up in normalization by a factor of 10
to make the figure less confusing.  The model curves are the best
fitting disk blackbody spectra.  We note the systematic positive
residuals in the ``very high state'' spectrum presents tenative
evidence for a hard tail in the highest luminosity systems, much like
their Galactic counterparts.}
\end{figure*}

This result is not particularly robust, since when we restrict the
summed spectra to those with globular cluster counterparts, both the
high/soft state and the very high state spectra fit well to a single
disk blackbody model with temperatures consistent with being equal.
Thus it is possible that one or more of the high flux sources without
an optical counterpart is a background active galaxy with a hard
spectrum and not a very high state black hole system in NGC 4472.
This issue merits further attention in larger samples of sources.  The
bulk of the sources should be sufficiently well separated for XMM to
observe them; XMM spectra could help determine whether there are
stronger power law tails in the systems above 4$\times10^{38}$
ergs/second, since its energy coverage between 8 and 15 keV will
provide good measurements in a region of the spectrum that should be
dominated by the power law tail.

\subsection{Comparison between red and blue cluster X-ray sources}

Given that there is a clear difference between the formation
efficiency of X-ray binaries in the red globular clusters and that in
the blue globular clusters (Paper I), it is possible that there is a
fundamental difference between the X-ray properties of the sources in
the red globular clusters and those in the blue globular clusters.  We
thus extract summed spectra for the red and blue globular clusters
separately and fit them using the same procedures as we used to
compare the globular cluster and field sources.

We find that the blue globular clusters clearly have harder spectra
than the red ones.  The best fit power law indices of the blue and red
systems are $1.02\pm0.27$ and $1.46\pm^{0.11}_{0.10}$ respectively,
while the best fitting disk blackbody inner temperatures are
$2.8\pm^{6.6}_{1.1}$ keV and $1.37\pm^{0.21}_{0.18}$ keV,
respectively.  For a bremstrahlung model, the red globular clusters
are best fit by a temperature of $10.7\pm^{8.7}_{3.7}$ keV, while the
lower limit for the bremstrahlung temperature of the blue clusters'
summed spectrum is 28 keV.  The red clusters X-ray sources dominate
the globular cluster system's X-ray source population, so it is not
surprising that the summed spectral properties of the red clusters are
consistent with the globular cluster sample as a whole and very
similar to those of the field sources.  The difference between the red
and blue clusters' integrated X-ray spectra are illustrated in Figure
2.

\begin{figure*}
\plotone{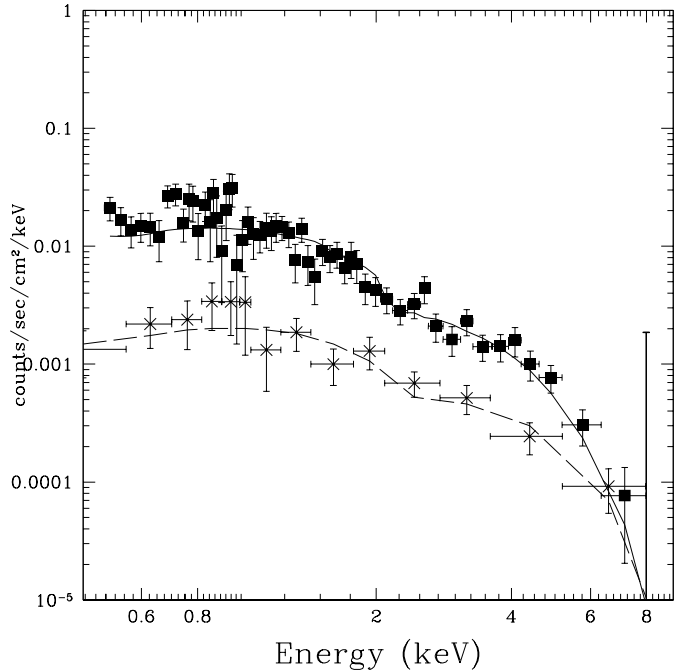}
\caption{The summed spectrum of the red globular clusters (filled
squares) and the blue globular clusters (X's), along with the best
fitting absorbed disk blackbody model. The red clusters' low mean disk
temperature can be seen by its steeper cutoff at high energies.}
\end{figure*}

While there is a statistically significant difference between the
X-ray spectral properties of the red and blue globular cluster X-ray
sources, we believe it is premature to ascribe this difference to the
metallicity.  There are only 7 X-ray sources in the blue globular
cluster systems, and the two brightest of these systems contribute
about 70\% of the total luminosity in the blue globular cluster X-ray
source sample.  The luminosity distribution of the 23 red globular
cluster X-ray sources appears less top heavy, but the two luminosity
functions cannot be distinguished with a KS test.  Because there are
so few X-ray sources in the blue globular cluster systems, a degree of
caution must be exercised in making broad statements about their
nature.  We thus believe only that it will be important to perform
similar analyses on the globular cluster systems of other galaxies in
order to determine whether the source spectra depend upon the globular
cluster's color.  Some credence to the notion that this result is
physically significant comes from the result of Irwin \& Bregman
(1999) that showed the most metal rich globular clusters in the local
group had the softest spectra in the soft X-ray band observed by
ROSAT.

\section{How are the LMXBs formed?}

A key result of this investigation and of Paper I is that the
properties of globular cluster LMXBs are generally consistent with
those of the field LMXBs.  There are several mechanisms for creating a
population LMXBs which can be found in the field population rather
than in the globular clusters of elliptical galaxies - (1) the
presence of a large, low duty cycle transient population (Piro \&
Bildsten 2001), (2) ejection of binaries formed in globular clusters
(e.g. Phinney \& Sigurdsson 1991), and (3) tidal disruption of
globular clusters, releasing systems formed in clusters into the field
(see e.g. Fall \& Zhang 2001; Vesperini 2000; Ostriker \& Gnedin
1997).  We will discuss how the observational results found here may
help to determine the relative importance of the different scenarios.
As we discuss the ``dominant'' formation mechanism here, we refer to
the mechanism for forming the {\it observed} sources; given a
sufficiently low duty cycle, there could exist many more quiescent
LMXBs than LMXBs in outburst, and those sources could constitute the
bulk of the LMXB population of a galaxy while at the same time
contibuting a small fraction of the observed sources.

\subsection{Spatial distributions and evidence for an ejection model}
The spatial distribution of the field and globular cluster LMXBs of
NGC~4472 are consistent with one another and with the optical globular
cluster distribution, but appear to be different from the galaxy's
field light profile (see Figure 4 of Paper I).  In fact, the globular
clusters show a shallower surface density than the integrated light
not only in NGC~4472, but in most elliptical galaxies (see e.g. Ashman
\& Zepf 1998).  The similarity between the LMXB and GC spatial
distributions (and the differences of these from the stellar light
profile) in NGC~4472 suggest that the bulk of the observed LMXBs are
formed in globular clusters.  If this result is confirmed in
observations of additional galaxies, it would indicate that mechanism
(1) above is relatively less important than mechanisms (2) and (3).
If mechanism (3) were the most important, then one would expect the
field LMXB distribution to be more centrally concentrated than the
current observed cluster distribution, as tidal destruction of
clusters is most effective in the center of the galaxy where the tidal
forces are strongest (Vesperini et al. 2002, in prep.). Mechanism (2)
predicts that the field LMXBs should have a slightly more spatially
extended distribution than the globular cluster distribution as a
whole, as the escape velocity from a globular cluster is about 10
km/sec and the lifetime of a bright persistent LMXB is about 300
million years or less, so the system will travel about 5 kiloparsecs,
or about 1 arcminute from its initial location at the distance of NGC
4472.  This distance will be large enough to make the initial globular
cluster associated with the LMXB impossible to determine (except in
the outer regions of the galaxy), but small enough that there will not
be a large population of LMXBs which have traveled very far from the
center of the galaxy.

The similar spectral properties of the field and cluster sources in
NGC~4472 provide additional evidence that the two populations have a
common origin.  At the present time, no work has been done to show
whether the long timescale transients should spend the same fraction
of their lives in the different spectral states as do the shorter
period transients and persistent sources, but it would seem to be an
unusual coincidence if they did.  The identical luminosity functions
and summed spectra of the two populations present a challenge to any
mechanism which posits a different origin for the cluster sources than
for the bulk of the field sources.  As noted above, the luminosity
functions present a stronger constraint on these grounds than do the
spectral properties.  Spectral properties are generally well-tied to a
source's fraction of the Eddington luminosity, so given identical
luminosity functions and identical proportions of black holes and
neutron stars, one would expect {\it a priori} identical summed
spectral properties.  The proportion of black holes and neutron stars
among the sources above $10^{37}$ ergs/sec may differ for transient
and persistent source, but this is presently not clear.  Hence, at the
present, the identical luminosity functions present stronger evidence
for the common origin of the bulk of the field and globular cluster
sources in NGC~4472.

It should be noted that mechanisms (1) and (3) may be at play in
addition to mechanism (2).  In fact, transient systems have been seen
in the field LMXB populations of elliptical galaxies (Kraft et
al. 2001), although it will not be clear without multi-epoch
observations of some elliptical galaxies what fraction of these
transients have long recurrence timescales and hence low duty cycles.
Some or all of them may be like the Galactic source Aql X-1, which is
a transient, but has a recurrence timescale of roughly one year and a
duty cycle of about 20\%.  Additionally, low duty cycle transients may
be present within the globular clusters as well as the field sources.

There are several reasons why, if ejections supply a substantial
fraction of the field sources, then these ejections are likely to be
caused by collisions and not by supernova kicks.  Firstly, binaries
immediately ejected by supernova kicks take no advantage of the
stellar interaction rates in the globular clusters - they should be
formed at the same rate in globular clusters as in the field, which
means that they should account for $\simlt$ 0.1\% of the field X-ray
binaries in their host galaxies.  Secondly, the binaries most likely
to be long period, low duty cycle transients when their secondary
stars evolve off the main sequence (Piro \& Bildsten 2001) are also
more likely to have small velocity kicks (Brandt \& Podsiadlowski
1995; BP95).  Thirdly, it has been suggested that black hole systems
tend to have systematically lower velocity kicks than neutron stars
(BP95), which means that if the ejected systems leave immediately
after becoming the supernova explosions, then the globular cluster
systems would be expected to have a substantially larger black hole
fraction than the field, while the observations show no signature of
this effect.

\subsection{Implications of the GC color-spectral index correlation}
If the correlation between globular cluster color and spectral index
of the X-ray sources is verified, possible causes might be a
dependence on metallicity of the relative proportion of black hole and
neutron star primaries in the X-ray binaries or of the mean binary
separation and hence mean accretion rate that causes different
fractions of the systems to be in different spectral states depending
on the metallicity.  It may be, for example, that the dependence of
binary fraction on globular cluster color is partly due to the fact
that the red globular clusters are younger than the blue ones, rather
than due to the fact that they are more metal rich, and that X-ray
binaries are preferentially formed early in the lifetimes of globular
clusters.  If the primary formation mechanism is the hardening of
existing binary systems by stellar interactions so that they can
undergo Roche lobe overflow, rather than the formation of new compact
object binary systems by tidal captures, then the depletion of a ``
reservoir'' of binary systems may cause the younger (red) clusters to
have more X-ray emitting systems than the older (blue) clusters.  If
this scenario is correct, then one would expect the blue globular
cluster X-ray binary sample to have a larger fraction of transient
sources, which would in turn possibly have different spectral
characteristics than the steady accretors.  At the present, no
definitive statement can be made on this topic.  Our future work will
include observations of systems such as NGC 4365, NGC 3115 and NGC
1399 where there is broader band optical data which can be used to
place stronger constraints on the red globular clusters' ages, and
theoretical investigations to determine whether the ``reservoir
depletion'' scenario outlined above is capable of quantitatively
reproducing the observations.  Additionally, multiple epoch
observations with Chandra can help determine whether the blue cluster
sources are more likely than the red cluster sources to be transient
systems and whether the field or globular clusters sources are more
likely to be transients.  At present, only one elliptical galaxy in
which the X-ray point source population can be well studied, NGC 5128,
has been observed twice and it was found that a substantial fraction
of its sources were transients (Kraft et al. 2001).

\subsection{Comparisons with other galaxies and possible evidence for an {\it in situ} population}

Through preliminary work we have done on the S0 galaxy NGC 3115 and
the elliptical galaxy NGC 4365, we have found that roughly 25\% and
35\%, respectively, of the X-ray binaries in these systems are
associated with globular clusters (the details of this work will
appear in a future paper).  This would seem to suggest that the
fraction of X-ray binaries in globular clusters varies with Hubble
type, with the highest fraction of globular cluster sources in the
earliest type galaxies.  There may additionally be some evidence for a
correlation between the fraction of X-ray binaries in globular
clusters and the fraction of total stellar mass in globular clusters,
as NGC 1399 has the highest values in the currently well-studied
sample for both these quantities and NGC 3115, the Milky Way and M31
have lower values for both these quantities (although recent star
formation could explain the large number of field LMXBs in the Milky
Way and M31).  NGC 4472 and NGC 4365 have intermediate values for both
quantities.

This correlation, if verified, would probably imply that a significant
fraction of the field sources are, in fact, generated in the field.
There would then be three populations of X-ray binaries - (a) the
globular cluster sources (b) the sources generated in globular
clusters but released into the field and (c) the population of field
X-ray binaries generated {\it in situ}.  Let us define the number of
globular cluster sources to be $N_{GC}$, the number of field sources
created in globular clusters to be $N_{FGC}$, and the number of field
sources preoduced {\it in situ} to be $N_{IS}$.  Furthermore, let us
define the ratio between the number of globular cluster X-ray sources
and the number of field sources produced in globular clusters to be
$\eta$, so that $N_{FGC}=\eta N_{GC}$.  We then find that:
\begin{equation}
\frac{N_{GC}}{N_{LMXB}} = \frac{N_{GC}}{N_{GC}(1+\eta)+N_{IS}},
\end{equation}
where $N_{LMXB}$ is the total number of LMXBs in the galaxy.  Thus the
fraction of sources in globular clusters may depend upon $\eta$,
$\frac{N_{GC}}{N_{IS}}$, or both.  It might be expected, for example,
that in the lenticular galaxy NGC 3115, which has the lowest estimated
value of $\frac{N_{GC}}{N_{LMXB}}$ among the early-type galaxies, that
$\eta$ could be especially large because more stellar interactions
within the globular clusters are caused by the greater tidal forces in
the disk-like galaxy.  Studies of a much larger sample of galaxies are
needed to separate the effects of $\eta$ and $\frac{N_{GC}}{N_{IS}}$.

We also find in NGC 4365 and NGC 3115 that the fraction of globular
clusters with LMXBs is about 4\%.  This fraction is also 4\% in
NGC~4472 (Paper I) and NGC 1399 (Angelini, Loewenstein \& Mushotzky
2001), and is 2-3\% in M31 (di Stefano et al. 2002; Barmby \& Huchra
2001) and 1-4\% (Liu, van Paradijs \& van den Heuvel 2001; Harris
1996) in the Milky Way with the large uncertainties coming from the
difficulty in counting the known transients.  Thus while this quantity
may have some effects on the relative fractions of LMXBs in globular
clusters in the spiral galaxies compared to the elliptical galaxies,
it seems to be fairly universal in the ellipticals and not to be a
major factor causing the wide range in the fraction of LMXBs in the
field versus the fraction in globular clusters.

\subsection{Discussion of selection effects}
There are a few causes for concern in applying the observational
constraints directly to the theoretical models.  There are signficant
selection effects in the X-rays that make our detected sample
incomplete.  The X-ray completeness may vary with radial distance from
the center of the galaxy due to both the degradation of the Chandra
point spread function as one moves off-axis and the variation of the
effective background from the diffuse gas as a function of
radius. Preliminary results seem to suggest that the two effects
balance one another for NGC 4472 (Kim \& Fabbiano 2002).  There will
also be similar selection effects in the optical (the faintest
clusters will be missed in the central regions of the galaxy because
of the contamination from the field starlight).  These should not be
too severe as it is likely that there are very few X-ray sources in
the faint globular clusters below the sensitivity limit of the HST
observations (see Paper I).  We {\it can} be fairly confident that the
spatial distribution and the luminosity function for the field and
cluster X-ray sources are consistent with one another, since the
selection effects should apply to the two X-ray populations equally.
The results of Kim \& Fabbiano (2002) suggest that spatial profile of
the X-ray sources is also unaffected by selection effects, so the
similarity between the spatial profiles of the optical globular
cluster sample and the X-ray sources is also likely to be a real
effect.  The final potential problem with gleaning results from these
data is that there are still relatively few sources.  With only 30
sources in globular clusters, and only 11 of those at luminosities
higher than the Eddington luminosity for a neutron star, comparisons
of luminosity functions, for example, are likely to suffer from small
number statistics.

\section{Very Massive Black Holes}

It has been recently noted that there may be relatively large
populations of very massive black holes (VMBHs) produced by supernova
explosions of population III stars (Schneider et al. 2002) since the
inefficiency of cooling of metal-free gas results in a top-heavy
initial mass function (Abel et al. 2000; Bromm et al. 2001), and the
supernova explosions of these heavy stars may produce black holes that
contain a very large fraction of the initial mass being locked up in
the final black hole (see e.g. Heger \& Woosley 2001).  It has further
been suggested that Bondi-Hoyle accretion from the interstellar medium
by the VMBHs could explain the bright non-nuclear X-ray source
population (Schneider et al. 2002).  On purely observational grounds,
we cannot rule out the hypothesis that isolated accreting VMBHs
represent some fraction of the X-ray sources, but we note that our
finding that 40\% of the X-ray sources are associated with globular
clusters indicates that the observations of these sources cannot be
used as a lower limit for the number of VMBHs, and that subsequent
lower limits on the redshift of star formation and on the fraction of
Population III stars that contribute to the metal enrichment of the
early universe (Schneider et al. 2002) must be revised.  The
similarities of all the X-ray properties of the globular cluster and
field X-ray sources would further suggest that VMBHs represent a small
fraction of the field sources.

Theoretical arguments also indicate that VMBHs are unlikely to make up
any of the observed X-ray sources in NGC 4472 and other elliptical
galaxies.  For NGC 4472 (and in fact, for most elliptical galaxies),
the Bondi-Hoyle accretion rate (Bondi 1952) is far too low to supply a
200 $M_\odot$ black hole with enough mass to be observable at the
detection threshold of $\sim 10^{37}$ ergs/sec.  The Bondi-Hoyle
accretion rates of the $\sim 10^9 M_\odot$ black holes at the centers
of elliptical galaxies are typically of order $1 M_\odot$/yr (di
Matteo et al. 2000), and given that the Bondi-Hoyle rate scales with
$M^2$ of the accreting object, the accretion rate for a $\sim200
M_\odot$ black hole should be $\sim 10^{-14} M_\odot$/yr, or
$7\times10^{11}$ g/sec, which would provide a luminosity of
$6\times10^{31}$ ergs/sec given the standard $0.1c^2$ efficiency, or a
substantially lower luminosity given the reduced radiative
efficiencies expected to be seen at very low accretion rate (see
e.g. Rees et al. 1982; Narayan \& Yi 1994; Blandford \& Begelman 1999;
di Matteo et al. 2000; Quataert \& Gruzinov 2000).  Because the
Bondi-Hoyle accretion rate also goes as $T^{-3/2}$, the neutral
interstellar medium may actually contribute more mass than the hot
X-ray emitting gas; for NGC 4472 (as in most elliptical galaxies), the
upper limit on the neutral hydrogen gas mass is a factor of $\sim$200
lower than the mass of the X-ray emitting gas (Kumar \& Thonnard
1983).  We can thus place a limit on the Bondi-Hoyle accretion rate of
the neutral hydrogen at $1\times10^{14}$ g/sec and hence on the
luminosity of $1\times10^{34}$ ergs/sec, so we still would not expect
to observe such sources.  Accretion onto isolated VMBHs thus may be of
some importance in spiral galaxies, where the interstellar medium may
be colder and denser (in molecular clouds, for example), and hence the
Bondi-Hoyle accretion rate may be higher, but cannot explain X-ray
sources in any low redshift elliptical galaxies except the ones most
rich in neutral hydrogen.  As VMBHs should be undetectable in NGC
4472, they may still contribute to the dark matter budget as suggested
by Schneider et al. (2002).

\section{The nuclear emission}

Three point sources are detected within 8'' of the nucleus.  The
brightest, source 71 in the X-ray list, is likely to be a very weak
active galactic nucleus, with a luminosity of about $10^{39}$
ergs/second.  Previous studies of these same data for NGC 4472 failed
to identify this source as a point source (Loewenstein et al. 2001),
most likely because these analyses used the CELLDETECT algorithm to
identify point sources rather than the more sensitive WAVDETECT
algorithm used in our work.  The 0.5-8 keV luminosity of our detected
source is very close to the 2-10 keV upper limit determined by
Loewenstein et al. (2001), so conclusions about the radiative
efficiency of this source are not significantly changed.  Observations
of weak radio lobes in NGC 4472 indicate a minimum energy content of
only $10^{54}$ ergs with a corresponding equipartition magnetic field
of $10^{-5}$ G, which gives a minimum particle lifetime of $\sim10^8$
years, and hence requires a minimum luminosity of only
$\sim3\times10^{38}$ ergs/sec (Ekers \& Kotanyi 1978).  Hence, it is
not surprising to find that the X-ray luminosity of the NGC 4472
nucleus is so low.  We find the position of the optical center to be
$\alpha=12:29:46.78$, $\delta=8:00:02.2$ in the USNO system, offset by
about 1'' arcsecond from the X-ray center, and within the rounding
errors of the radio core position (Ekers \& Kotanyi 1978).  We find an
offset of about 2'' with regards to the SIMBAD optical center
position, but our optical center should be far more reliable and we
find good agreement with the position of the center reported by NED.

The two other point sources are collinear with the presumed nucleus
and the major axes of their source ellipses as detected by WAVDETECT
are approximately parallel to the line connecting the sources.
Furthermore, the line connecting the sources is perpendicular to the
weak core-dominated radio jets seen in previous observations (Ekers \&
Kotanyi 1978).  Recently, equatorial radio emission has been seen
simultaneously with jet emission in the Galactic source SS 433 (Paragi
et al. 1998,1999a,1999b; Blundell et al. 2001), and it has been
suggested that at least a fraction of the X-ray emission seen from
this source might come from the equatorial outflow if the equatorial
radio emission is produced by bremstrahlung processes (Blundell et
al. 2001).  Strong disk winds are predicted at very high accretion
rates (in sources such as the likely super-Eddington SS 433) or low
accretion rates in sources such as NGC 4472 by theoretical work (see
e.g. Blandford \& Begelman 1999), and may be responsible for the
production of such ``ruffs'' and the very low radiative efficiency of
the AGN in NGC 4472.  The radio observations initially used to detect
the jet in NGC 4472 had very poor North-South angular resolution and
would not have been capable of resolving a ruff in this source.  Deep,
high resolution radio images are needed to test whether the equatorial
X-ray sources are associated with a disk wind or the alignment of the
sources on an axis perpendicular to the radio jet is a coincidence.

\section{Conclusions}

Source lists for the optical and X-ray sources in NGC 4472 are
presented.  The X-ray properties of the sources associated with
globular clusters are consistent with those of the sources without
optical counterparts.  The mean X-ray spectrum of the sources in the
blue globular clusters is harder than that in the red globular
clusters but this result may be spurious because there are so few
X-ray sources in the blue globular clusters.  There is marginal
evidence for a hard tail in the sources whose luminosities indicate
they are in the very high state, and no evidence for such a tail in
the sources whose luminosities indicate they are in the high/soft
state.  We show that the current observational evidence suggests
tenatively that the bulk of the field sources were ejected from
globular clusters.  On the other hand, a tentative correlation between
the fraction of LMXBs in globular clusters and the fraction of stellar
mass in globular clusters across the four well-studied elliptical
galaxies suggests that there may also be a field population generated
{\it in situ}. We discuss future observations which may help test this
picture.  We show that accreting isolated very massive black holes
cannot produce observable X-ray sources in NGC 4472 or most other
elliptical galaxies.  There is possible evidence for an equatorial
``ruff'' near the galactic nucleus which may be associated with a disk
wind.

\section{Acknowledgments}

We gratefully thank Kathy Rhode for providing us with her large field
of view optical image which was essential for this study to tie
together accurately the various HST images to the Chandra image and
all the satellite images to the USNO system.  We thank Lars Bildsten
for a critical review of this manuscript.  We thank the anonymous
referee for several useful suggestion which helped improve the clarity
and the technical accuracy of this paper.  This research has made use
of the NASA/IPAC Extragalactic Database (NED) which is operated by the
Jet Propulsion Laboratory, California Institute of Technology, under
contract with the National Aeronautics and Space Administration.  This
research has made use of the SIMBAD database, operated at CDS,
Strasbourg, France.  SEZ and AK gratefully acknowledge support from
NASA via the LTSA grant NAG5-11319.

\begin{table}
\begin{center}
\caption{The X-ray Source List}
\begin{tabular}{lllllllllllll}
\tableline\tableline
ID&IAU Name&$\Gamma$&$\sigma_\Gamma+$&$\sigma_\Gamma-$&$F_x(0.5-2.0)$&$F_x(2.0-8.0)$&$L_x$&$\chi^2$&$\nu$&Good Region&HST Region&OID\\
\tableline
   1&CXOMKZ J1229314+080055&2.00&    &    &8.07E-16&8.50E-16&5.19E+37& 0.0& 3&Y&N\\
   2&CXOMKZ J1229331+080218&1.66&0.44&0.41&2.69E-15&4.53E-15&2.25E+38& 3.6& 4&Y&N\\
   3&CXOMKZ J1229345+080032&1.18&0.17&0.16&7.05E-15&2.27E-14&9.15E+38&14.9& 8&Y&N\\
   4&CXOMKZ J1229350+080049&1.20&0.45&0.37&1.83E-15&5.78E-15&2.35E+38& 2.8& 3&Y&N\\
   5&CXOMKZ J1229359+080054&2.00&    &    &6.56E-16&6.91E-16&4.22E+37& 0.0& 2&Y&N\\
   6&CXOMKZ J1229368+080311&1.02&0.56&0.49&2.15E-15&8.65E-15&3.32E+38& 0.3& 3&Y&N\\
   7&CXOMKZ J1229369+080311&1.09&0.55&0.47&2.17E-15&7.96E-15&3.12E+38& 0.0& 3&Y&N\\
   8&CXOMKZ J1229378+080046&2.00&    &    &6.53E-16&6.88E-16&4.20E+37& 0.0& 2&Y&N\\
   9&CXOMKZ J1229378+075846&1.28&0.56&0.51&4.76E-15&1.35E-14&5.64E+38& 0.2& 3&Y&N\\
  10&CXOMKZ J1229381+075838&2.00&    &    &2.28E-15&2.40E-15&1.46E+38& 0.0& 2&Y&N\\
  11&CXOMKZ J1229382+080235&2.00&    &    &1.60E-15&1.69E-15&1.03E+38& 0.0& 3&Y&N\\
  12&CXOMKZ J1229382+080013&2.00&    &    &5.87E-16&6.19E-16&3.78E+37& 0.0& 2&Y&N\\
  13&CXOMKZ J1229383+080030&1.20&0.54&0.51&2.01E-15&6.37E-15&2.59E+38& 0.2& 3&Y&N\\
  14&CXOMKZ J1229392+075920&2.00&    &    &1.33E-15&1.40E-15&8.54E+37& 1.2& 3&Y&N\\
  15&CXOMKZ J1229401+075927&1.47&0.33&0.29&3.45E-15&7.48E-15&3.38E+38& 0.3& 4&Y&N\\
  16&CXOMKZ J1229401+075944&2.00&    &    &5.34E-16&5.62E-16&3.43E+37& 0.0& 2&Y&N\\
  17&CXOMKZ J1229402+075944&2.00&    &    &1.18E-15&1.25E-15&7.63E+37& 0.6& 3&Y&N\\
  18&CXOMKZ J1229402+075829&2.00&    &    &1.77E-15&1.87E-15&1.14E+38& 0.0& 3&Y&Y  & 4\\
  19&CXOMKZ J1229403+080241&2.00&    &    &3.22E-16&3.39E-16&2.07E+37& 0.0& 2&Y&N\\
  20&CXOMKZ J1229404+080225&2.00&    &    &9.27E-16&9.76E-16&5.96E+37& 0.0& 3&Y&N\\
\tableline
\end{tabular}
\tablecomments{The complete version of this table is in the electronic
edition of the Journal.  The printed edition contains only a sample.
The units are ergs/sec/cm$^2$ for the fluxes and ergs/sec for the
luminosities. The ``good region'' is the region outside the central 8
arcseconds.  Numbers in the OID column refer to the sources' optical
counterparrts' ID numbers in Table 3.}
\end{center}
\end{table}

\begin{table}
\begin{center}
\caption{The Optical Source List}
\begin{tabular}{llllllllllllllllll}
\tableline\tableline
ID&RA&&&Dec&&&$V$&$\sigma_V$&$I$&$\sigma_I$&$V-I$&$\sigma_{V-I}$&$R_{HL}$&$\sigma_R$&CFOV&GC&XID\\
\tableline
1&12&29&40.0&+8& 3&29.8&22.13&0.01&21.07&0.01&1.07&0.01& 0.0&1.5&Y&Y&\\
2&12&29&40.2&+7&58& 1.5&23.16&0.02&21.85&0.01&1.31&0.03& 0.9&1.5&N&Y&\\
3&12&29&40.2&+7&57&59.9&25.75&0.17&24.51&0.13&1.24&0.22&12.8&1.2&N&Y&\\
4&12&29&40.2&+7&58&30.0&22.77&0.02&21.80&0.02&0.97&0.02& 2.4&0.9&Y&Y&18\\
5&12&29&40.4&+7&57&55.0&23.57&0.03&22.71&0.03&0.86&0.04& 1.4&0.6&N&Y&\\
\tableline
\end{tabular}
\tablecomments{The complete version of this table is in the electronic
edition of the Journal.  The printed edition contains only a sample.
The half-light radii are measured in arcseconds, a ``Y'' in the CFOV
column indicates that a source is in the Chandra field of view, a
``Y'' in the GC column indicates a source is a good globular cluster
candidate and a number in the XID column refers to the X-ray
counterpart's ID number in Table 2.}
\end{center}
\end{table}

\begin{table}
\begin{center}
\caption{X-Ray/HST Matches}
\begin{tabular}{llllllllllllll}
\tableline\tableline
{\bf XID}&{\bf IAU Name}&{\bf $\Gamma$}&{\bf F(.5-2 keV)} &{\bf F(2-8 keV)} & {\bf Lum} & $\chi^2$ & $\nu$ &{\bf OID} &{\bf Position} & $V$ & $I$& $V-I$& $R_{hl}$ \\ 
\tableline
  18& CXOMKZ J1229402+075829&   2.00(fixed)& 1.77$\times10^{-15}$& 1.87$\times10^{-15}$& 1.14$\times10^{38}$&   0.0&   3&    4& 12 29 40.2 +7 58 30.0& 22.77$\pm{ 0.02}$& 21.80$\pm{ 0.02}$& 0.97$\pm{ 0.02}$&  2.4$\pm{ 0.9}$\\
  22& CXOMKZ J1229410+080003&   2.00(fixed)& 9.35$\times10^{-16}$& 9.84$\times10^{-16}$& 6.01$\times10^{37}$&   1.4&   3&   29& 12 29 41.0 +8  0  3.7& 21.88$\pm{ 0.02}$& 20.61$\pm{ 0.02}$& 1.26$\pm{ 0.03}$&  1.4$\pm{ 0.9}$\\
  27& CXOMKZ J1229416+080059&   1.56$\pm^{   3.76}_{   1.78}$& 9.18$\times10^{-16}$& 1.76$\times10^{-15}$& 8.29$\times10^{37}$&   0.4&   3&   68& 12 29 41.6 +8  0 59.4& 23.11$\pm{ 0.06}$& 22.26$\pm{ 0.06}$& 0.85$\pm{ 0.08}$&  0.0$\pm{ 1.8}$\\
  28& CXOMKZ J1229416+080015&   0.96$\pm^{   0.36}_{   0.37}$& 2.87$\times10^{-15}$& 1.26$\times10^{-14}$& 4.75$\times10^{38}$&   2.1&   4&   70& 12 29 41.6 +8  0 14.8& 22.88$\pm{ 0.04}$& 21.56$\pm{ 0.04}$& 1.32$\pm{ 0.06}$&  3.2$\pm{ 1.1}$\\
  30& CXOMKZ J1229416+080045&   2.00(fixed)& 7.93$\times10^{-16}$& 8.34$\times10^{-16}$& 5.09$\times10^{37}$&   1.3&   3&   75& 12 29 41.6 +8  0 45.1& 21.64$\pm{ 0.02}$& 20.37$\pm{ 0.02}$& 1.27$\pm{ 0.02}$&  0.6$\pm{ 0.9}$\\
  31& CXOMKZ J1229416+080049&   2.00(fixed)& 5.36$\times10^{-16}$& 5.64$\times10^{-16}$& 3.44$\times10^{37}$&   0.0&   2&   76& 12 29 41.7 +8  0 49.8& 21.99$\pm{ 0.02}$& 20.96$\pm{ 0.02}$& 1.03$\pm{ 0.03}$&  0.0$\pm{ 1.2}$\\
  34& CXOMKZ J1229423+080008&   1.26$\pm^{   0.15}_{   0.15}$& 9.56$\times10^{-15}$& 2.79$\times10^{-14}$& 1.16$\times10^{39}$&  12.9&  11&  137& 12 29 42.4 +8  0  8.4& 20.57$\pm{ 0.01}$& 19.30$\pm{ 0.01}$& 1.27$\pm{ 0.01}$&  2.3$\pm{ 0.9}$\\
  38& CXOMKZ J1229430+080040&   2.00(fixed)& 1.36$\times10^{-15}$& 1.43$\times10^{-15}$& 8.73$\times10^{37}$&   5.0&   3&  209& 12 29 43.1 +8  0 40.2& 21.11$\pm{ 0.01}$& 20.10$\pm{ 0.01}$& 1.01$\pm{ 0.02}$&  1.2$\pm{ 0.9}$\\
  44& CXOMKZ J1229438+080041&   1.43$\pm^{   0.52}_{   0.48}$& 2.01$\times10^{-15}$& 4.65$\times10^{-15}$& 2.06$\times10^{38}$&   0.5&   3&  274& 12 29 43.8 +8  0 41.5& 21.20$\pm{ 0.02}$& 19.90$\pm{ 0.01}$& 1.29$\pm{ 0.02}$&  1.0$\pm{ 0.8}$\\
  45& CXOMKZ J1229438+075957&   2.00(fixed)& 9.67$\times10^{-16}$& 1.02$\times10^{-15}$& 6.23$\times10^{37}$&   0.4&   3&  280& 12 29 43.8 +7 59 57.3& 21.12$\pm{ 0.01}$& 20.02$\pm{ 0.01}$& 1.11$\pm{ 0.02}$&  1.5$\pm{ 1.3}$\\
  48& CXOMKZ J1229440+075952&   1.29$\pm^{   0.45}_{   0.38}$& 2.41$\times10^{-15}$& 6.73$\times10^{-15}$& 2.82$\times10^{38}$&   4.2&   4&  299& 12 29 44.0 +7 59 52.0& 23.19$\pm{ 0.07}$& 21.92$\pm{ 0.05}$& 1.27$\pm{ 0.09}$&  1.1$\pm{ 1.3}$\\
  50& CXOMKZ J1229442+080042&   2.09$\pm^{   1.01}_{   0.59}$& 1.48$\times10^{-15}$& 1.39$\times10^{-15}$& 9.05$\times10^{37}$&   0.3&   3&  317& 12 29 44.2 +8  0 42.5& 20.86$\pm{ 0.01}$& 19.62$\pm{ 0.01}$& 1.24$\pm{ 0.02}$&  0.0$\pm{ 1.1}$\\
  52& CXOMKZ J1229450+075950&   1.52$\pm^{   0.27}_{   0.25}$& 4.62$\times10^{-15}$& 9.38$\times10^{-15}$& 4.33$\times10^{38}$&   1.3&   6&  408& 12 29 45.1 +7 59 50.9& 21.83$\pm{ 0.03}$& 20.51$\pm{ 0.03}$& 1.32$\pm{ 0.04}$&  0.0$\pm{ 1.4}$\\
  61& CXOMKZ J1229457+075743&   2.00(fixed)& 1.72$\times10^{-15}$& 1.81$\times10^{-15}$& 1.10$\times10^{38}$&   2.3&   3&  509& 12 29 45.8 +7 57 44.1& 21.60$\pm{ 0.01}$& 20.29$\pm{ 0.01}$& 1.31$\pm{ 0.01}$&  4.0$\pm{ 1.0}$\\
  64& CXOMKZ J1229461+075915&   1.84$\pm^{   0.26}_{   0.23}$& 6.22$\times10^{-15}$& 8.12$\times10^{-15}$& 4.46$\times10^{38}$&   3.8&   8&  543& 12 29 46.1 +7 59 15.9& 23.83$\pm{ 0.06}$& 22.48$\pm{ 0.05}$& 1.35$\pm{ 0.08}$&  6.4$\pm{ 0.7}$\\
  65& CXOMKZ J1229463+075949&   1.12$\pm^{   1.02}_{   4.12}$& 9.20$\times10^{-16}$& 3.26$\times10^{-15}$& 1.29$\times10^{38}$&   2.1&   6&  566& 12 29 46.3 +7 59 49.3& 21.38$\pm{ 0.01}$& 20.24$\pm{ 0.01}$& 1.14$\pm{ 0.02}$&  2.7$\pm{ 0.2}$\\
  66& CXOMKZ J1229464+080127&   2.00(fixed)& 9.40$\times10^{-16}$& 9.90$\times10^{-16}$& 6.04$\times10^{37}$&   2.9&   3&  571& 12 29 46.4 +8  1 27.3& 24.17$\pm{ 0.08}$& 23.12$\pm{ 0.07}$& 1.05$\pm{ 0.11}$&  3.8$\pm{ 0.8}$\\
  75& CXOMKZ J1229470+075847&   1.73$\pm^{   1.01}_{   0.69}$& 1.34$\times10^{-15}$& 2.04$\times10^{-15}$& 1.05$\times10^{38}$&   0.0&   3&  663& 12 29 47.0 +7 58 47.4& 21.79$\pm{ 0.01}$& 20.55$\pm{ 0.01}$& 1.24$\pm{ 0.01}$&  1.4$\pm{ 0.7}$\\
  79& CXOMKZ J1229477+075926&   1.07(bad fit)& 1.68$\times10^{-15}$& 6.32$\times10^{-15}$& 2.46$\times10^{38}$&   9.9&   4&  748& 12 29 47.8 +7 59 26.3& 22.63$\pm{ 0.03}$& 21.36$\pm{ 0.03}$& 1.26$\pm{ 0.04}$&  0.8$\pm{ 0.9}$\\
  82& CXOMKZ J1229479+075919&   2.00(fixed)& 1.27$\times10^{-15}$& 1.33$\times10^{-15}$& 8.12$\times10^{37}$&   0.1&   3&  764& 12 29 47.9 +7 59 19.6& 21.33$\pm{ 0.01}$& 20.13$\pm{ 0.01}$& 1.20$\pm{ 0.01}$&  0.0$\pm{ 1.2}$\\
  90& CXOMKZ J1229491+08 027&   2.00(fixed)& 7.27$\times10^{-16}$& 7.66$\times10^{-16}$& 4.68$\times10^{37}$&   5.2&   3&  861& 12 29 49.1 +8  0 27.9& 22.90$\pm{ 0.03}$& 21.73$\pm{ 0.03}$& 1.17$\pm{ 0.04}$&  1.4$\pm{ 0.8}$\\
  92& CXOMKZ J1229493+075753&   0.97$\pm^{   0.38}_{   0.39}$& 2.95$\times10^{-15}$& 1.28$\times10^{-14}$& 4.84$\times10^{38}$&   2.5&   4&  882& 12 29 49.4 +7 57 54.0& 21.00$\pm{ 0.01}$& 20.00$\pm{ 0.01}$& 1.00$\pm{ 0.01}$&  4.2$\pm{ 0.9}$\\
  93& CXOMKZ J1229495+075925&   2.00(fixed)& 9.34$\times10^{-16}$& 9.83$\times10^{-16}$& 6.00$\times10^{37}$&   0.0&   3&  909& 12 29 49.6 +7 59 25.7& 21.59$\pm{ 0.01}$& 20.30$\pm{ 0.01}$& 1.29$\pm{ 0.01}$&  1.8$\pm{ 0.8}$\\
  95& CXOMKZ J1229501+075944&   1.89$\pm^{   0.78}_{   0.53}$& 1.59$\times10^{-15}$& 1.94$\times10^{-15}$& 1.10$\times10^{38}$&   2.0&   4&  966& 12 29 50.1 +7 59 44.2& 21.41$\pm{ 0.01}$& 20.28$\pm{ 0.01}$& 1.13$\pm{ 0.01}$&  1.7$\pm{ 0.8}$\\
  99& CXOMKZ J1229504+080014&   1.59$\pm^{   0.25}_{   0.24}$& 4.79$\times10^{-15}$& 8.89$\times10^{-15}$& 4.25$\times10^{38}$&   5.1&   6&  985& 12 29 50.4 +8  0 14.2& 21.26$\pm{ 0.01}$& 19.96$\pm{ 0.01}$& 1.30$\pm{ 0.01}$&  1.6$\pm{ 1.2}$\\
 102& CXOMKZ J1229509+08 010&   1.35$\pm^{   0.32}_{   0.30}$& 4.29$\times10^{-15}$& 1.10$\times10^{-14}$& 4.72$\times10^{38}$&   3.9&   5& 1008& 12 29 50.9 +8  0 10.4& 23.61$\pm{ 0.04}$& 22.45$\pm{ 0.05}$& 1.16$\pm{ 0.06}$&  0.0$\pm{ 0.8}$\\
 106& CXOMKZ J1229513+075917&   2.00(fixed)& 9.30$\times10^{-16}$& 9.80$\times10^{-16}$& 5.98$\times10^{37}$&   1.1&   3& 1030& 12 29 51.4 +7 59 17.7& 21.75$\pm{ 0.01}$& 20.57$\pm{ 0.01}$& 1.18$\pm{ 0.01}$&  0.4$\pm{ 0.9}$\\
 111& CXOMKZ J1229517+075960&   2.48$\pm^{   3.54}_{   0.88}$& 1.27$\times10^{-15}$& 6.96$\times10^{-16}$& 6.26$\times10^{37}$&   0.0&   3& 1041& 12 29 51.7 +8  0  0.5& 20.39$\pm{ 0.00}$& 19.14$\pm{ 0.00}$& 1.25$\pm{ 0.01}$&  1.0$\pm{ 0.6}$\\
 113& CXOMKZ J1229525+080002&   2.00(fixed)& 7.67$\times10^{-16}$& 8.08$\times10^{-16}$& 4.93$\times10^{37}$&   0.9&   3& 1066& 12 29 52.5 +8  0  2.4& 23.56$\pm{ 0.04}$& 22.35$\pm{ 0.03}$& 1.21$\pm{ 0.05}$&  0.0$\pm{ 1.2}$\\
 121& CXOMKZ J1229533+075952&   1.07$\pm^{   0.51}_{   0.55}$& 1.78$\times10^{-15}$& 6.70$\times10^{-15}$& 2.61$\times10^{38}$&   1.0&   4& 1086& 12 29 53.4 +7 59 52.4& 21.99$\pm{ 0.01}$& 20.98$\pm{ 0.01}$& 1.01$\pm{ 0.02}$&  1.3$\pm{ 1.2}$\\
\tableline
\end{tabular}
\end{center}
\end{table}

\end{document}